\documentstyle[prb,aps,epsf]{revtex}

\begin{document}

\draft
\title{Numerical renormalization-group study of spin correlations in
one-dimensional random spin chains}
\author{T.\ Hikihara}
\address{Graduate School of Science and Technology, Kobe University,
Rokkodai, Kobe 657-8501, Japan}
\author{A.\ Furusaki and M.\ Sigrist}
\address{Yukawa Institute for Theoretical Physics, Kyoto University,
Sakyo-ku, Kyoto 606-8502, Japan}
\date{\today}
\maketitle
\begin{abstract}
We calculate the ground-state two-spin correlation functions of
spin-$\frac{1}{2}$ quantum Heisenberg chains with random exchange couplings
using the real-space renormalization group scheme.
We extend the conventional scheme to take account of the contribution of
local higher multiplet excitations in each decimation step.
This extended scheme can provide highly accurate numerical data
for large systems.
The random average of staggered spin correlations of the chains with
random antiferromagnetic (AF) couplings shows algebraic decay like $1/r^2$,
which verifies the Fisher's analytic results.
For chains with random ferromagnetic (FM) and AF couplings,
the random average of generalized staggered correlations is found to
decay more slowly than a power-law, in the form close to $1/\ln(r)$.
The difference between the distribution functions of the spin correlations
of the random AF chains and of the random FM-AF chains is also discussed.
\end{abstract}

\section{Introduction}
One-dimensional (1D) quantum spin systems have attracted much
attention over the decades.
This is not only because these systems have been a
good testing ground for various theoretical techniques and
approximations but also because they exhibit a wealth of fascinating
phenomena in their ground states and low-lying excitations.
These include quasi long-range-order (LRO), topological order
and ground-state phase transitions, which are all purely quantum
effects due to the low-dimensionality of the systems.
Among these quantum phenomena, the effects of randomness on quantum
spin systems have been studied intensively by many groups.
These studies revealed the appearance of various exotic phases,
which are realized neither in regular quantum systems
nor classical random systems.
The interplay of randomness and quantum fluctuations plays an
essential role in these phases.
Systems which are gapless in the absence of randomness are
unstable against weak randomness,\cite{MDH,Fisher} while gapped
systems, e.g.~integer spin chains in the Haldane phase and dimerized
chains, are comparatively robust.\cite{Hyman,Fabrizio,Monthus}

A simplest model of the 1D random spin-$\frac{1}{2}$ Heisenberg spin
systems is described by the Hamiltonian of the form
\begin{equation}
H = \sum_{i=1}^{L-1} J_i \vec{s}_i \cdot \vec{s}_{i+1},
\label{eq:Ham}
\end{equation}
where $\vec{s}_i$ are $S = \frac{1}{2}$ spin operators and
the exchange coupling constants $J_i$ are distributed randomly
according to a probability distribution $P(J_i)$.
There are several quasi-1D systems which provide the realizations of
the model Hamiltonian (\ref{eq:Ham}).
To our knowledge, the first example of such systems belongs to
the class of the organic charge-transfer salts tetracyanoquinodimethanide
(TCNQ).\cite{TCNQ}
The low-temperature magnetic properties of these systems are
successfully described by the model Hamiltonian (\ref{eq:Ham}) with
random antiferromagnetic (AF) coupling,\cite{GT}
in which the couplings $J_i$ are restricted to take positive random
values.
A more recently studied system\cite{pmJ} is
Sr$_3$CuPt$_{1-x}$Ir$_x$O$_6$.
While the pure compounds Sr$_3$CuPtO$_6$ and Sr$_3$CuIrO$_6$
represent, respectively, AF and ferromagnetic (FM) spin chains,
Sr$_3$CuPt$_{1-x}$Ir$_x$O$_6$ contains both AF and FM couplings.
The fraction of FM bonds is simply related to the concentration $x$ of
Ir.
These compounds are modeled adequately by the Hamiltonian (\ref{eq:Ham})
with $P(J_i) = p \delta(J_i + J) + (1-p) \delta(J_i-J) $, where $p$ is
the probability of FM bonds.
The experimental data of the susceptibility of the compounds were in
fact explained successfully by a theory based on the Hamiltonian
(\ref{eq:Ham}) with the above probability distribution of FM and AF
bonds.\cite{Furu}
The model (\ref{eq:Ham}) is also realized in the low-temperature regime
of randomly depleted 1D spin-$\frac{1}{2}$ even-leg
ladders.\cite{2legE}
In these systems, effective $S = \frac{1}{2}$ spins are induced in the
vicinity of each depleted
site.\cite{Imada,Fukuyama,SigFuru,Martins,Mikeska,Miyazaki,Greven}
If the density of depleted spins is sufficiently small, the induced
effective spins are the only degrees of freedom which are active in
the low-energy limit.
The strength of the residual interaction between effective spins
depends on the distance between the effective spins, and
the coupling can be either antiferromagnetic or ferromagnetic,
depending on whether or not two spins are depleted from the same
sublattice.
Thus, the low-energy physics of the randomly depleted
spin-$\frac{1}{2}$ even-leg ladders is described as a random spin
chain with both AF and FM exchange couplings.\cite{SigFuru}

For 1D random spin chains containing only AF couplings, rather
complete theoretical understanding on their low-energy properties has
been achieved.
One of the most powerful techniques to study such systems is
the real-space renormalization group (RSRG) method
introduced by Ma, Dasgupta, and Hu.\cite{MDH}
The basic idea of this method is iterative decimation of spin degrees
of freedom by integrating out the strongest bonds in the chain
successively.
For the random AF chains this procedure keeps the form of the
Hamiltonian (\ref{eq:Ham}) and renormalizes the probability
distribution $P(J_i)$.
Fisher has given a solution to this RG equation of $P(J_i)$ that
becomes exact in the low-energy limit and shown that any normalizable
initial distribution $P(J_i)$ flows to a single universal fixed point
distribution.\cite{Fisher}
The ground state of the chain which belongs to the fixed point is
characterized by the ^^ ^^ random singlet phase," where each spin
forms a singlet with another spin which can be arbitrarily far away.
From the intuitive picture of the random singlet phase, Fisher has
also shown that the random average of the two-spin correlation
function in this phase decays algebraically like $1/r^2$ with $r$
being the distance between the two spins.
The main contribution to the average comes from the rare events that
two spins separated by the distance $r$ form a singlet pair.
The $1/r^2$ power-law decay has been verified by a numerical
calculation for the random $XX$ spin-$\frac{1}{2}$
chains,\cite{Henelius} which can be mapped to a disordered system of
noninteracting fermions.

The random spin chains containing both AF and FM couplings have also
been studied theoretically for the last five years, from which
a qualitative picture of the low-energy physics of such
random FM-AF chains has emerged.
Westerberg {\it et al.}~\cite{West} have adapted the RSRG scheme
and shown through extensive numerical simulations that the
distribution $P(J_i)$ is renormalized to a single universal
fixed-point distribution unless the initial distribution is highly
singular around $J_i = 0$.
The chain at this fixed point can be viewed as an ensemble of weakly
interacting large effective spins whose average size
$\bar{S} \propto T^{-\kappa}$ with $\kappa \approx 0.22$ for
temperature $T \to 0$.
These large effective spins are generated as a result of decimations
of two spins coupled via a strong FM coupling into a larger spin.
The results of the RSRG simulations are supported by a recent
calculation by Frischmuth {\it et al}.,~\cite{Fri} who have used the
continuous-time quantum Monte Carlo loop algorithm.
In contrast to the success in the quantitative calculations on the
thermodynamic properties, less is known about the spin correlations in
the random FM-AF chains.
Since the average size of the effective spins becomes very large in
the low-energy (long-distance) limit, one can expect that the system
should be close to a classical spin chain that can show the LRO of
the generalized staggered spin component.
One might also argue, however, that there should be no LRO in 1D
quantum spin chains.
It is therefore an interesting open question how the correlation
function of the generalized staggered moment (whose definition will be
given in Sec.\ III) behaves in the ground state.
The purpose of this paper is in fact to present results of our
extensive numerical calculation of the two-spin correlation function.
We find that it decays very slowly with the form close to $1/\ln(r)$,
which is consistent with the naive argument that the ground state is
extremely long-range correlated, but not really long-range ordered.

The outline of this paper is as follows.
Section \ref{sec:conventionalRSRG} is devoted to a brief review of the
RSRG scheme of the generalized version which is applicable
to both random AF and FM-AF case.
Using the Wigner-Eckart theorem,\cite{W-E} we simplify the algorithm
to allow for calculating the two-spin correlation function
between original $S=\frac{1}{2}$ spins.
In section \ref{sec:extension}, in order to achieve higher accuracy,
we extended the RSRG scheme to take into account the contributions of
local excitations to the ground state of the whole system.
We perform these conventional and extended RSRG algorithm numerically
on both random AF and FM-AF chains to calculate the ^^ ^^ staggered"
spin correlations on their ground states.
The results on the random AF case are shown in section \ref{sec:AF}.
In section \ref{sec:FMAF}, we show the results on the random FM-AF
case.
Analyzing the data obtained with the extended RSRG method (which can
be applied to larger systems than the DMRG),
we conclude that the mean correlations on the random FM-AF case
decay very slowly with logarithmic dependence on $r$.
We also discuss the distribution functions of the logarithm of
the correlation functions in section \ref{sec:Dis}.
We find that, in the random AF case, the rare spin pairs,
which are strongly correlated, dominate the mean correlation
while such rare events are not essential in the random FM-AF case.
Finally, our results are summarized in section \ref{sec:conclusion}.

\section{The RSRG algorithm}

Ma, Dasgupta and Hu introduced a RSRG procedure to investigate
the low-temperature properties of random AF spin chains.\cite{MDH}
The method has been generalized to the random FM-AF case
by Westerberg {\it et al}.\cite{West}
In this section we explain our extension of this scheme
to calculate the ground-state two-spin correlation functions.
We begin with a brief review of the RSRG method for the random FM-AF
case.

\subsection{Conventional RSRG}
\label{sec:conventionalRSRG}

Let us consider a random FM-AF spin-$\frac{1}{2}$ chain described by
the Hamiltonian (\ref{eq:Ham}).
The basic strategy of RSRG is to decimate spin degrees of freedom
by combining two spins connected via a strong bond into one
effective spin.
Consequently, the system is described in terms of effective spins of
various sizes coupled by random exchange interactions,
although the original Hamiltonian (\ref{eq:Ham}) consists of only
$S=1/2$ spins.
We accordingly treat the effective Hamiltonian,
\begin{equation}
H = \sum_l J_l \vec{S}_l \cdot \vec{S}_{l+1},
\label{eq:Heff}
\end{equation}
where both the coupling $J_l$ and the size of the effective spins $S_l$
are random.
We call the $S=1/2$ spins $\vec{s}_i$ appearing in the Hamiltonian
(\ref{eq:Ham}) ^^ ^^ original" spins in the following to distinguish
them from the effective spins $\vec{S}_l$.

We define $\Delta_l$ as the energy gap between the ground-state multiplet
and the first excited multiplet of the corresponding bond Hamiltonian
$H_l = J_l \vec{S}_l \cdot \vec{S}_{l+1}$,
\begin{eqnarray*}
\Delta_l = \left\{
 \begin{array}{@{\,}ll}
 |J_l|(S_l+S_{l+1}) & (J_l < 0) \\
 J_l(|S_l-S_{l+1}|+1) & (J_l > 0)
 \end{array}
 \right.
\end{eqnarray*}
and focus on the bond with the largest gap $\Delta_l$ in the chain.
The terms in the effective Hamiltonian (\ref{eq:Heff}) which involve
the effective spins $\vec{S}_l$ and $\vec{S}_{l+1}$ are
\begin{equation}
H' = H'_0 + H'_1 \label{eq:H'}
\end{equation}
where
\begin{eqnarray}
H'_0 & = & J_l \vec{S}_l \cdot \vec{S}_{l+1} \label{eq:H'0} ,\\
H'_1 & = & J_{l-1} \vec{S}_{l-1} \cdot \vec{S}_l
           + J_{l+1} \vec{S}_{l+1} \cdot \vec{S}_{l+2} \label{eq:H'1} .
\end{eqnarray}
If $\Delta_l$ is much larger than the gaps of the neighboring bonds,
$\Delta_{l-1}$ and $\Delta_{l+1}$, the spins $\vec{S}_l$ and
$\vec{S}_{l+1}$,
to a good approximation, are frozen into the ground-state multiplet
of the local Hamiltonian $H'_0$.
We, therefore, replace the block composed of $\vec{S}_l$ and $\vec{S}_{l+1}$
by the single effective spin $\vec{S}$.
The Wigner-Eckart theorem~\cite{W-E} then implies that both $\vec{S}_l$
and $\vec{S}_{l+1}$ are proportional to $\vec{S}$:
\begin{eqnarray}
\vec{S}_l & = & \alpha \vec{S}, \nonumber \\
\vec{S}_{l+1} & = & \beta \vec{S}, \label{eq:WE}
\end{eqnarray}
where $\alpha$ and $\beta$ can be obtained from
the Clebsch-Gordan coefficients.
Substituting Eq.~(\ref{eq:WE}) into
the four-spin Hamiltonian (\ref{eq:H'}), we obtain,
apart from a constant term coming from $H'_0$,
\begin{equation}
\widetilde{H} = \tilde{J}_{l-1} \vec{S}_{l-1} \cdot \vec{S}
         + \tilde{J}_{l+1} \vec{S} \cdot \vec{S}_{l+2}, \label{eq:Hnew}
\end{equation}
where
\begin{eqnarray*}
\tilde{J}_{l-1} = \alpha J_{l-1} , \\
\tilde{J}_{l+1} = \beta J_{l+1} .
\end{eqnarray*}
The case where $J_l$ is antiferromagnetic with $S_l = S_{l+1}$
needs a special treatment.
In this case the two spins $\vec{S}_l$ and $\vec{S}_{l+1}$ form a
singlet, and accordingly, we remove both spins from the effective
Hamiltonian.
Between the spins $\vec{S}_{l-1}$ and $\vec{S}_{l+2}$,
a coupling is generated through a second-order process that
virtually breaks the singlet of $\vec{S}_l$ and $\vec{S}_{l+1}$.
We obtain
\begin{equation}
\widetilde{H} = \tilde{J} \vec{S}_{l-1} \cdot \vec{S}_{l+2},
\label{eq:Hnew'}
\end{equation}
where
\begin{eqnarray*}
\tilde{J} = \frac{2 J_{l-1} J_{l+1}}{3 J_l} S_l (S_l + 1) .
\end{eqnarray*}

\begin{figure}
\epsfxsize=11cm
\epsffile{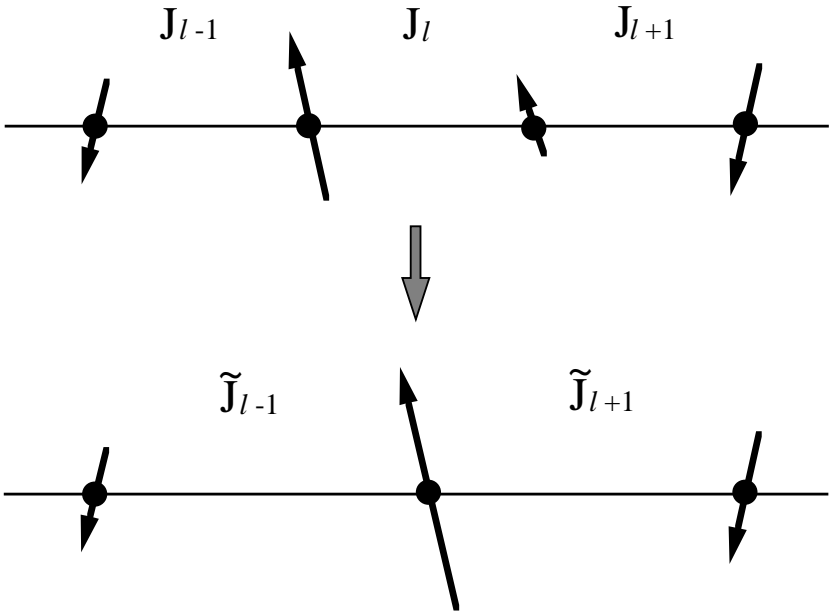}
   \caption{Schematic picture of renormalization procedure in the
conventional
            RSRG.}
  \label{fig:conRSRG}
\end{figure}

By replacing the four-spin Hamiltonian (\ref{eq:H'})
with $\widetilde{H}$ [Eq.~(\ref{eq:Hnew}) or Eq.~(\ref{eq:Hnew'})]
in the Hamiltonian of the whole system, we obtain a new effective
Hamiltonian (See Fig.~\ref{fig:conRSRG}).
We note that this procedure preserves the form of the effective
Hamiltonian (\ref{eq:Heff}) but changes the distributions of the
exchange couplings, $J_l$,
and the size of effective spins, $S_l$.
We repeat this procedure of integrating out the strongest bonds
in a chain successively until the distribution functions for $J_l$ and
$S_l$ converge to a scaling form.
This RG flow has been investigated extensively for various initial
distributions including both AF and FM-AF case.~\cite{MDH,West}
In particular, for the random AF $S=1/2$ chains Fisher has solved the
RG equation exactly.~\cite{Fisher}

It is also possible to calculate the correlation functions
between original spins within the scheme described above.
Here we make use of the fact that the original spin operator
$\vec{s}_i$ is, at every step in the RSRG procedure, proportional to
the effective spin $\vec{S}_l$ to which it belongs.
We can keep track of the coefficient for each original spin operator
by applying Eq.~(\ref{eq:WE}) at each step.
At the step where the effective spins $\vec{S}_l$ and
$\vec{S}_{l+1}$ is added into $\vec{S}$, we calculate the correlations
between $\vec{s}_i$ and $\vec{s}_j$ in the ground state of
the bond Hamiltonian $H'_0$,
\begin{equation}
\langle \vec{s}_i \cdot \vec{s}_j \rangle
= \alpha_i \alpha_j \langle \vec{S}_l \cdot \vec{S}_{l+1} \rangle,
\label{<s_is_j>g}
\end{equation}
where $\langle \cdots \rangle$ represents the expectation value
in the ground state; $\vec{s}_i$ and $\vec{s}_j$ belong to $\vec{S}_l$ and
$\vec{S}_{l+1}$, respectively; $\alpha_i$ and $\alpha_j$ are the
proportionality coefficients for each spin.

\subsection{Extension of the RSRG scheme}
\label{sec:extension}

As seen in the previous subsection, the conventional RSRG procedure
consists of ^^ ^^ diagonalizing the bond Hamiltonian with the
largest gap" and ^^ ^^ projecting the low-energy states onto
the lowest multiplet."
This means that we completely neglect the contribution of
the excited multiplets of the local Hamiltonian (\ref{eq:H'0})
to the ground state of the whole system.
This approximation is valid only if the energy gap of $H'_0$
is much larger than the ones of the neighboring bonds.
Fisher's solution for the random AF $S=1/2$ chain becomes
asymptotically exact since this condition is satisfied near the fixed
point of the RG flow.

However, the condition is often not satisfied, in particular, in the
early stage of the RG, unless the initial distribution of energy gap
is very broad.
As a result, we have no choice but to cut off the contributions of
local excitations.
This is a poor approximation that has a serious effect especially on
the calculation of the expectation values of microscopic operators
such as the original spins.
The ground-state correlation functions between original spins
calculated via the conventional RSRG in fact deviate largely from
those obtained from the DMRG method and the extended RSRG algorithm
described below (See Fig.~\ref{fig:AFC} and Fig.~\ref{fig:FMC} in
section \ref{sec:results}).

One possible prescription to avoid this error is
^^ ^^ to keep the local multiplet excitations" at each step.
Of course, if we keep all eigenstates of the bond Hamiltonian $H'_0$
at every step of the RG, the calculation on $H'_0$ in the final step
is equivalent to the exact diagonalization on the Hamiltonian of the
whole system.
In practice what we should do is to extend the RSRG algorithm under
the policy that we keep as many states as we can store in computer
memories.
In the original RSRG scheme a segment of original spins are combined
and represented by a single effective spin.
In the extended scheme we keep more states than the lowest multiplet
and call the segment a ^^ block.'
Each block consists of several original spins and is represented by
^^ block states,' which are the $m$-lowest eigenstates
of the block Hamiltonian.
Since we have to keep or discard all states of a multiplet to ensure
the SU(2) symmetry, the actual number of kept states for block $l$ is
$m^\ast_l \le m$.
An original spin operator in block $l$ is represented by
$m^\ast_l \times m^\ast_l$ matrix on the set of the block states
accordingly.

\begin{figure}
\epsfxsize=11cm
\epsffile{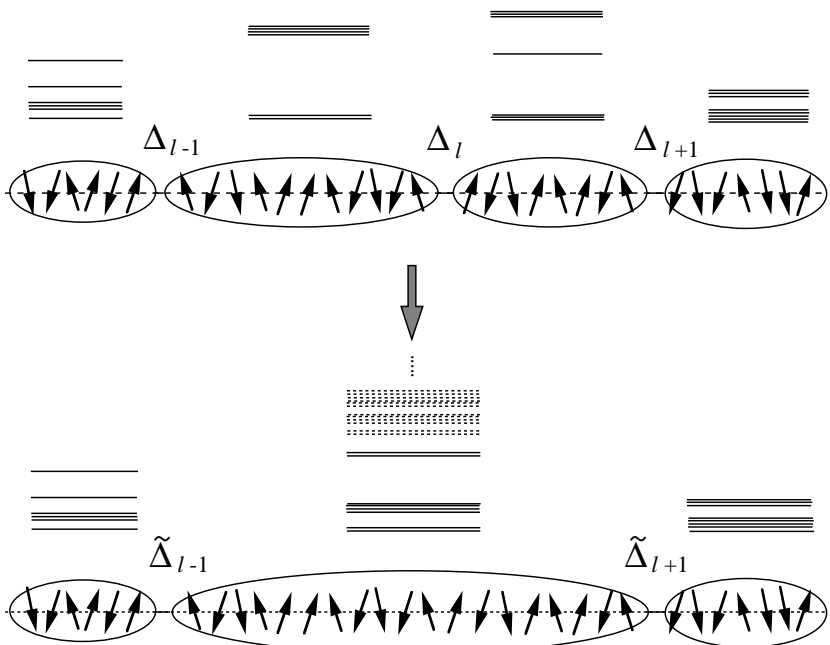}
   \caption{Schematic picture of renormalization procedure in the
            extended RSRG. The dashed and solid lines represent
            eigenstates of each block Hamiltonian which are discarded
            and kept, respectively.}
  \label{fig:extRSRG}
\end{figure}

Let us consider the effective Hamiltonian
\begin{eqnarray}
H & = & \sum_l H^{(B)}_l + \sum_l H_{l,l+1} \\
H_{l,l+1} & = & \tilde{J}_l \vec{s}^{\,(R)}_l \cdot
\vec{s}^{\,(L)}_{l+1},
\end{eqnarray}
where $H^{(B)}_l$ is a block Hamiltonian of the $l$th block,
diagonal in the block states; $H_{l,l+1}$ is a coupling
Hamiltonian between the $l$th and $(l+1)$th blocks;
$\vec{s}^{(R)}_l$ and $\vec{s}^{(L)}_l$ are original spin
operators on the right and left edge of the $l$th block, respectively;
$\tilde{J}_l$ is a coupling between $\vec{s}^{(R)}_l$ and
$\vec{s}^{(L)}_{l+1}$.
In the extended RSRG scheme we renormalize the two-block Hamiltonian
\begin{equation}
H'_{l,l+1} = H^{(B)}_l + H_{l,l+1} + H^{(B)}_{l+1}
\label{eq:2BH}
\end{equation}
with the largest gap $\Delta_l$ into one block Hamiltonian
(see Fig.~\ref{fig:extRSRG}).
Here we define $\Delta_l$ as the energy difference between the highest
energy in the eigen-multiplets of $H'_{l,l+1}$ which will be kept
and the lowest energy in the multiplets which will be discarded
after the decimation.
The basic scheme of the extended RSRG is the same as the conventional
RSRG except the changes described above.
The algorithm is summarized as follows:
\def\labelenumi{(\theenumi)}
\def\theenumi{\roman{enumi}}
\begin{enumerate}
\item Focus on the bond with the largest gap $\Delta_l$.
Construct the two-block Hamiltonian (\ref{eq:2BH}) of the bond.
\item Diagonalize the two-block Hamiltonian to find a set of
eigenvalues and eigenstates.
At this point, we can calculate expectation values of various
operators in the two blocks, such as the two-spin correlation
functions between the original spins, using the ground state of the
two-block Hamiltonian.
\item Discard all but the lowest $m^\ast_l (\le m)$ eigenstates in the
block Hamiltonian.
\item Express operators, such as the block Hamiltonian itself and the
original spin operators in the new block, in terms of the
new block states.
\item Rewrite the coupling Hamiltonians between the new block and its
neighboring blocks in terms of the new $\vec{s}^{(L)}$ and
$\vec{s}^{(R)}$.
Diagonalize them to update the distribution of the energy gap $\Delta$.
\item Return to the step ({\rm i}).
\end{enumerate}
We obtain all two-spin correlation functions by repeating this
procedure until the whole chain is finally represented by one block.

\section{Numerical Results}
\label{sec:results}

Using both the conventional RSRG and the extended RSRG with various
values of $m$,
we calculated the correlation functions
$\langle \vec{s}_i \cdot \vec{s}_j \rangle$ in the ground state of
open chains for both the random AF case and the random FM-AF case.
The maximum size of the chains used for the conventional and the
extended RSRG simulations is $L = 100000$ and $L = 1000$, respectively.
In the FM-AF case, the random average of the spin correlation,
$\langle \vec{s}_i \cdot \vec{s}_j \rangle$, always decays
exponentially\cite{West} and is not an appropriate quantity to
characterize the spin correlation because the sign of
$\langle \vec{s}_i \cdot \vec{s}_j \rangle$
can be either positive or negative depending on the number of AF bonds
between $\vec{s}_i$ and $\vec{s}_j$.
Instead we introduce the ^^ ^^ generalized staggered" correlation function
\begin{equation}
C(|i-j|) = \eta_{ij}^{} \langle \vec{s}_i \cdot \vec{s}_j \rangle,
\end{equation}
where $\eta_{ij} = \prod_{k=i}^{j-1} {\rm sgn}\left( -J_k \right)$
for $j > i$.
For random AF case, $C(r)$ is the usual staggered spin correlation.
We take the random average of $C(r)$ and $\ln C(r)$, which represents
the mean correlations and the logarithm of the typical correlations,
respectively.
Note that it is impossible to take the random average of
$\ln C(r)$ numerically in the conventional RSRG algorithm, within
which $C(r)=0$ for two spins that do not form a singlet pair.
To check the results of the RSRG methods, we also calculated the
correlation functions on $L=100$ open chains using the DMRG
method~\cite{White} with the improved algorithm proposed by
White.~\cite{impDMRG}
The number of kept states in the DMRG calculation was up to 100 and 200
for the random AF and FM-AF case, respectively.
In both cases, the mean and typical correlations calculated by
the extended RSRG are in excellent agreement with those by DMRG (see below).
We note that the systems we have treated are much bigger than those in
the earlier work by Hida,\cite{Hida} in which the DMRG method is
applied for the FM-AF chains.

\subsection{Random AF case}
\label{sec:AF}

As a typical probability distribution of the random AF case,
we choose the box-type bond distribution $P(J_i)$,
\begin{equation}
P(J_i) = \left\{
 \begin{array}{@{\,}ll}
 \frac{1}{J_0} & ( 0 \le J_i \le J_0 ) \\
 0             & ( {\rm otherwise} )
 \end{array}
 \right.
\label{P(J);AF}
\end{equation}
where the cutoff, $J_0$, is taken as energy unit.
For the random AF case, it is known that any normalizable initial
distribution flows to a single universal fixed point.~\cite{Fisher}
Hence, the results we obtained for the initial distribution
(\ref{P(J);AF}) are generic.
The number of sample chains we used for each method are shown in Table
\ref{tb1}.

Before discussing our numerical results, we briefly comment on the
$m$-dependence of the data of the extended RSRG.
As noted in the last section, the extended RSRG can provide
more accurate results as the number of kept states, $m$, increases.
In the random AF case, the ground-state multiplet of a block
Hamiltonian is either singlet or doublet, depending on the number of
original spins belonging to the block.
The degeneracy of each low-lying excited multiplet is, therefore, always
small.
As a result, we can keep considerably many multiplets even if
$m$ is rather small.
We estimate the $m$-dependence of the results of the extended RSRG
from the numerical data with $m=10, 20, 30$.

\begin{figure}
\epsfxsize=11cm
\epsffile{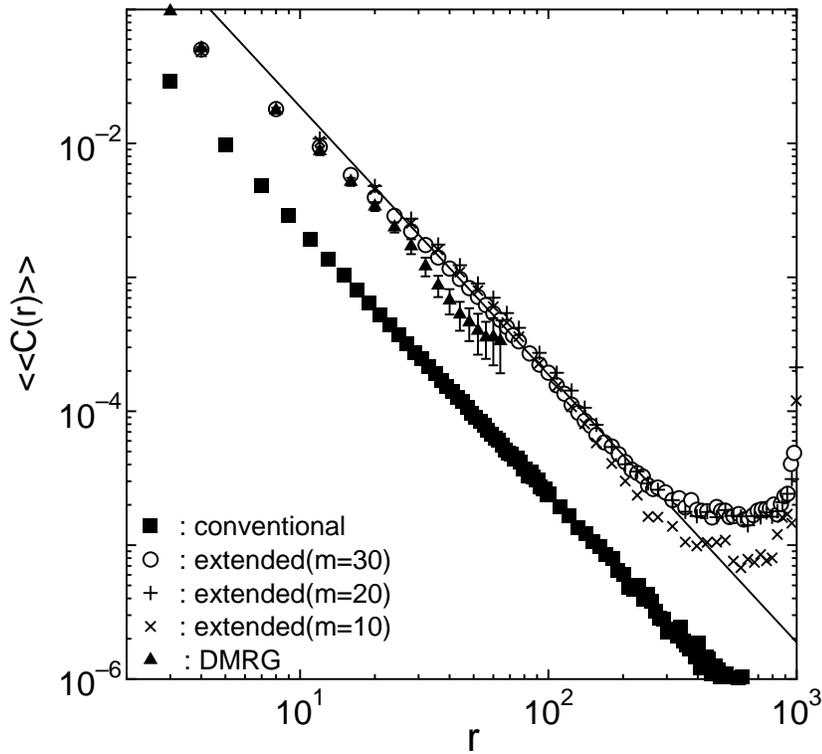}
   \caption{The mean correlation $\langle \langle C(r) \rangle
            \rangle$ of random AF chains as a function of $r$. The
            solid line corresponds to the $r^{-2}$ decay.
            The statistical errors of the data of the extended RSRG
            for $r < 300$ and of the conventional RSRG
            is smaller than the size of symbols.}
  \label{fig:AFC}
\end{figure}

\begin{figure}
\epsfxsize=11cm
\epsffile{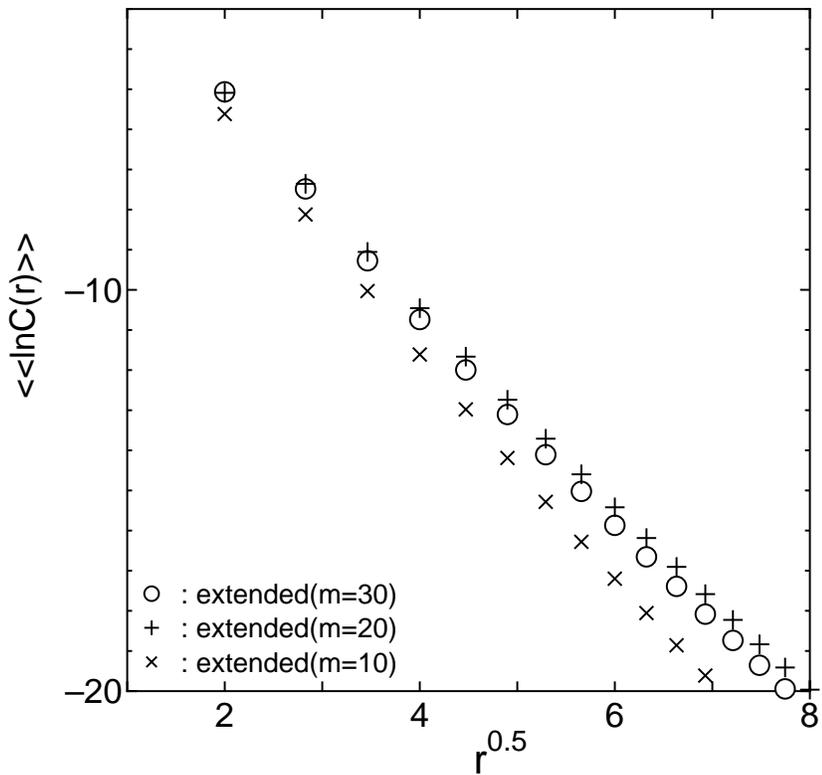}
   \caption{$\langle \langle \ln C(r) \rangle \rangle$ versus
            $r^{0.5}$ for the random AF case.
            The statistical errors of the data
            are smaller than the size of symbols.}
  \label{fig:AFlnC}
\end{figure}

Figures \ref{fig:AFC} and \ref{fig:AFlnC} show, respectively, the
numerical data of the mean correlations, $\langle \langle C(r) \rangle
\rangle$, and the average of the logarithm of correlations,
$\langle \langle \ln C(r) \rangle \rangle$,
where $\langle \langle \cdots \rangle \rangle$ represents random average.
It is clear that the data of the extended RSRG converge already at
$m=30$, and we can regard those data with $m=30$ as those in
the limit $m\to\infty$.
In Fig.~\ref{fig:AFC} the data of the extended RSRG with $m=30$ are
in good agreement with those of DMRG for $r\lesssim30$, where the DMRG
data are considered to be exact and free from finite-size effects.
On the other hand, the conventional RSRG largely underestimates the
correlations, but its data are on a line parallel to the data of the
extended RSRG in the log-log plot.
We conclude from these observations that the results of the extended
RSRG are quantitatively reliable, whereas the conventional RSRG
can be used to estimate the exponent of the power-law decay.
Encouraged by the agreement between the extended RSRG and the DMRG
data, we anticipate that the extended RSRG provide quantitatively
reliable data even for $r>30$ where the reliable DMRG data are not
available.
From the data of the extended RSRG with $m = 30$ for $r\lesssim300$,
where the data are expected not to be hampered by finite-size effects,
we estimate the asymptotic form of the average correlations to be
\begin{equation}
\langle \langle C(r) \rangle \rangle \sim r^{-2}. \label{eq:FS1}
\end{equation}
For the average of $\ln C(r)$, we also rely on the data obtained from
the extended RSRG scheme.
Figure \ref{fig:AFlnC} gives
\begin{equation}
\langle \langle \ln C(r) \rangle \rangle \sim -r^{0.5}. \label{eq:FS2}
\end{equation}
Both results (\ref{eq:FS1}) and (\ref{eq:FS2}) agree with Fisher's
theory~\cite{Fisher} and can be considered as a numerical
verification of his solution on the mean and typical
correlations for the Heisenberg case.
For the random $XX$ chains\cite{Henelius} and for a related model
of the random transverse-field Ising model,\cite{Young} the
power-law behavior (\ref{eq:FS1}) and (\ref{eq:FS2}) is observed
numerically.
To our knowledge, Figs.~\ref{fig:AFC} and \ref{fig:AFlnC} are
first numerical results confirming Fisher's theory for the
Heisenberg case.

\subsection{Random FM-AF case}
\label{sec:FMAF}

Westerberg {\it et al}.~\cite{West} showed that the RG trajectories of the
random FM-AF chains flow towards a single universal fixed point under
the conventional RSRG procedure, unless the initial distribution of
couplings is more singular than
$P(J_i) \sim |J_i|^{-y_c}$, $y_c \sim 0.7$.
In this section we investigate the spin correlations at this fixed
point with the extended RSRG scheme.
For this purpose we assume the box-type bond distribution $P(J_i)$,
\begin{equation}
P(J_i) = \left\{
 \begin{array}{@{\,}ll}
 \frac{1}{2J_0} & ( -J_0 \le J_i \le J_0 ), \\
 0             & ( {\rm otherwise} ),
 \end{array}
 \right.
\label{eq:P}
\end{equation}
with $J_0$ as energy unit, as a representative of distributions with no
singularity at $J_i=0$.
We expect that our results obtained for the initial distribution
(\ref{eq:P}) should capture the universal behavior for the random
FM-AF chains which are in the basin of the universal fixed point found
in Ref.~\onlinecite{West}.
We have numerically calculated spin correlations using the
conventional RSRG, the extended RSRG with $m = 30, 40, 50, 60$ and the
DMRG method.

In the random FM-AF chains the degeneracy of the lowest multiplet of a
block becomes larger on average as the size of the block grows.
At the point when the degeneracy exceeds the number of kept states,
$m$, determined beforehand in the algorithm, the extended RSRG scheme
breaks down because we need to keep all the degenerate states in the
lowest multiplet to preserve the SU(2) spin symmetry.
In fact, we could complete the extended RSRG procedure without
exceeding the limit of the number of kept states $m=30$
($m=60$) for 30\% (75\%) of the sample chains ($L=1000$).
We then used only those samples for which the RSRG could be completed
to take the random average.
Although this sorting out of sample chains may lead to
a systematic underestimate on the average values,
we believe that we can correct it by carefully checking
the $m$-dependence of the data.
The number of sample chains we used to take the random average for
each RG scheme is shown in Table \ref{tb2}.

The numerical results for the mean correlations,
$\langle \langle C(r) \rangle \rangle$,
are shown in Fig.~\ref{fig:FMC} in a log-log plot.
It is clear that the data of the extended RSRG have almost converged with
$m=60$ for $r < 500$, where the data are expected to be free from the
effect of the open boundaries.
The data with $m =60$ are also in excellent
agreement with the data of DMRG.
We, therefore, regard the results of the mean correlation with $m = 60$
as essentially converged.
We notice that the curve in the log-log plot are bent upward,
indicating that the mean correlations $\langle\langle
C(r)\rangle\rangle$ decay more slowly than a power-law.
To analyze the $r$-dependence of the mean correlations,
we plot the inverse of $\langle \langle C(r) \rangle \rangle$
as a function of $\ln r$ in the inset of Fig.~\ref{fig:FMC}.
We find that the data lie on a straight line in this plot, from which
we speculate that the mean correlations decay with the logarithmic form,
\begin{equation}
\langle \langle C(r) \rangle \rangle \sim
 \frac{a}{\ln \left(r / r_0 \right)},
\label{eq:log}
\end{equation}
where $a$ and $r_0$ are constants.
We note, however, that Eq.~(\ref{eq:log}) is not the only form that
can account for the numerical data.\cite{stretched}
In any case the mean correlations show very weak $r$-dependence,
probably through the form of $\ln(r/r_0)$, certainly different from
power-law.

Figure \ref{fig:FMlnC} shows the numerical results of
$\langle \langle \ln C(r) \rangle \rangle$.
The data of the extended RSRG with $m = 60$ exhibit a similar behavior
to the logarithm of the mean correlations;
The typical correlation
$\exp \left[ \langle \langle \ln C(r) \rangle \rangle \right]$
decays again more slowly than a power-law.
This leads us to plot the inverse of the typical correlations
as a function of $\ln r$ (see the inset of Fig.~\ref{fig:FMlnC}).
Although the curve at $m = 60$  seems almost linear for $50 < r < 400$,
we cannot determine the $r$-dependence of the typical correlations
from the figure due to the slow convergence of the data with increasing $m$.
This slow convergence arises from the fact that the numerical estimate
of $\ln C(r)$ is very sensitive to small fluctuations of $C(r)$
especially when the value of $C(r)$ is extremely small.
Calculations with much larger values of $m$ would be needed to obtain
the data accurate enough to determine the $r$-dependence of
$\langle \langle \ln C(r) \rangle \rangle$.

\subsection{Distribution of the correlation functions}
\label{sec:Dis}

In order to make a distinction between characteristics of ground-state
correlation functions in the random AF chains and in the random FM-AF
chains, we analyze the distributions of the logarithm of the
correlations,
$D(x;r)$, where $x = \ln C(r)$.
Henelius and Girvin\cite{Henelius} have shown numerically that for
large $r$ the distribution function of $XX$ chains with random AF
couplings scales to a fixed-point distribution of the form
\begin{equation}
D(x;r) = f(r) F\left( x/g(r) \right) \label{eq:scl}
\end{equation}
with
\begin{eqnarray}
&& f(r)g(r) = 1, \label{eq:scl1} \\
&& g(r) = c\left|\langle \langle \ln C(r) \rangle \rangle\right|,
\label{eq:scl2}
\end{eqnarray}
where $c$ is a positive constant.
The scaling function $F(x/g(r))$ satisfies the normalization condition
$\int F(y)dy=1$.
We will demonstrate that the distribution $D(x;r)$ in random Heisenberg
chains also exhibits the scaling behavior Eq.~(\ref{eq:scl}) for both
the random AF and the random FM-AF case.

Figure~\ref{fig:ASC} shows the data of $D(x;r)$ for the random AF case
obtained using the extended RSRG with $m = 30$.
According to Eqs.~(\ref{eq:FS2}) and (\ref{eq:scl2}),
we can set $g(r) = r^{0.5}$ and $f(r) = r^{-0.5}$.
The data points (circles and squares) in Fig.~\ref{fig:ASC} collapse
on a single curve, indicating that the scaling (\ref{eq:scl})
applies.\cite{note}
In Fig.~\ref{fig:ASC} we also plot
\begin{equation}
W(x;r)=\frac{e^xD(x;r)}{\langle \langle C(r) \rangle \rangle},
\label{W(x;r)}
\end{equation}
which measures the contribution to the mean value
of the correlation $\langle\langle C(r)\rangle\rangle$.
Although the curves are rather rough due to a statistical error,
it is clear that $W(x;r)$ has a considerable weight in the range where
$D(x;r)$ is very small.
This means that a very few spin pairs that have much stronger
correlations than typical ones give dominant contribution to the mean
correlation $\langle\langle C(r)\rangle\rangle$.
We also find that, as $r$ increases, the region of $x/g(r)$ where
$W(x;r)$ is large moves towards $x/g(r)=0$.
Indeed this is the behavior expected from Eq.~(\ref{eq:FS1});
the value of $x/g(r)$ where $W(x;r)$ takes a large weight should change
as $\ln \langle \langle C(r) \rangle \rangle /g(r) \sim \ln r /
r^{0.5}\to0$ as $r\to\infty$.
Thus we regard the results shown in Fig.~\ref{fig:ASC} as a further
support for the random-singlet picture of the ground state of the
random AF Heisenberg chains, where each spin forms a singlet pair with
another spin that can be arbitrarily far away.
The mean correlation $\langle\langle C(r)\rangle\rangle$ is dominated
by the contribution from the rare case in which two spins separated by
distance $r$ form a spin singlet.

\begin{figure}
\epsfxsize=11cm
\epsffile{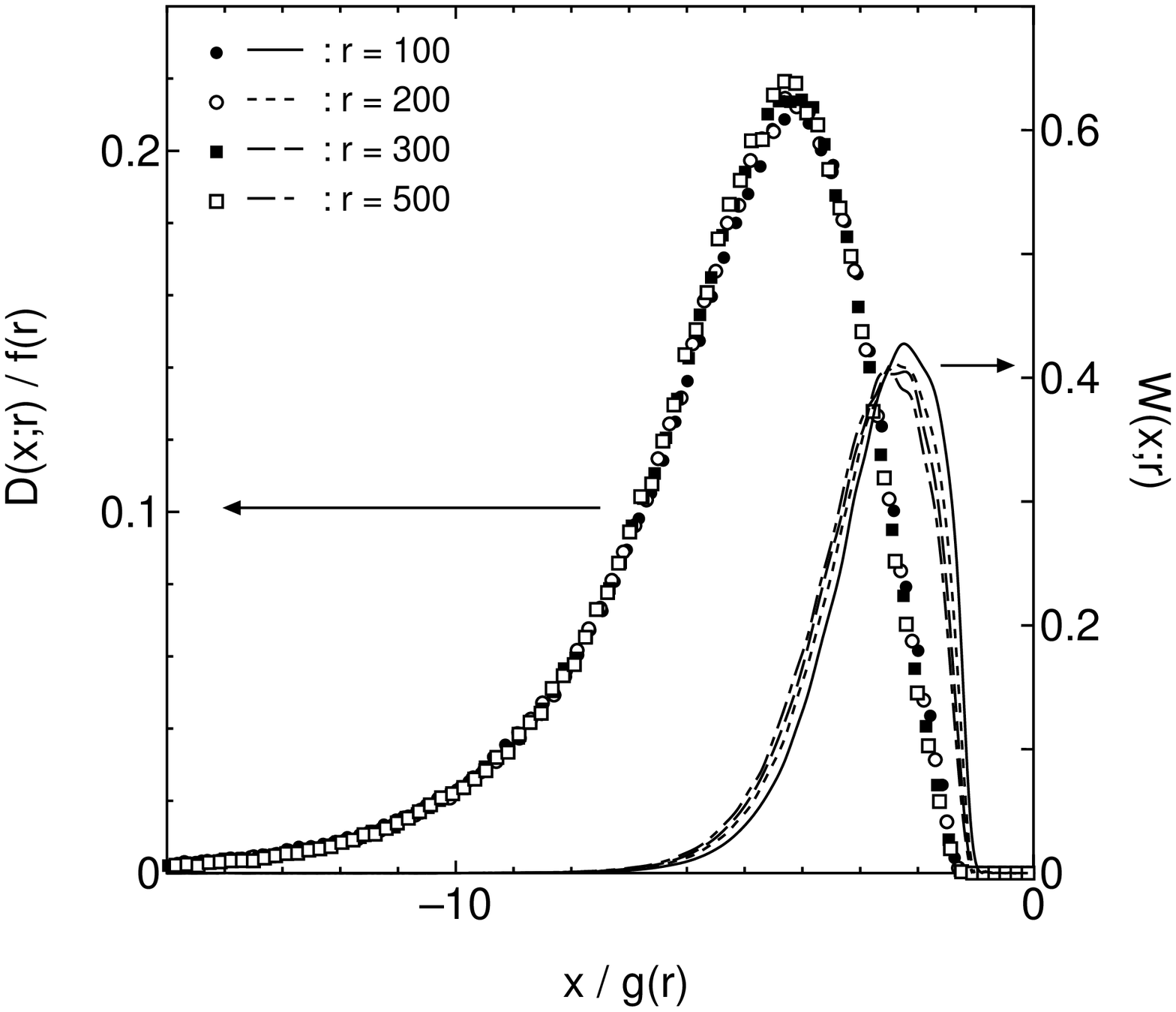}
   \caption{Probability distributions of the correlation functions
    in the random FM-AF case for $r = 100,200,300,500$.
    $W(x;r)$ is also plotted for $r = 100,200,300,500$ with a solid,
    dotted, dashed, dot-dashed curve, respectively.}
  \label{fig:FSC}
\end{figure}

Figure \ref{fig:FSC} shows the numerical data of $D(x;r)$ of the
extended RSRG with $m = 60$ for the random FM-AF chains.
Here we set $g(r) = \langle \langle \ln C(r) \rangle \rangle /
\langle \langle \ln C(r=200) \rangle \rangle $ and $f(r) = 1/g(r)$.
The data points (circles and squares) for $100 < r < 500$, where we
may ignore the boundary effect, clearly lie on a single scaling curve.
In this range of $r$, however, $g(r)$ changes by several percent only,
from $0.95$ to $1.04$.
To establish the scaling behavior for wide range of $g(r)$, the
calculations on much (exponentially) larger systems might be
necessary.
Such large-scale calculations are obviously unfeasible with computers
available at present, and thus we can only conclude that our results
shown in Fig.~\ref{fig:FSC} are consistent with the scaling hypothesis
(\ref{eq:scl}).
In the following discussion, we regard $D(x;r)/f(r)$ in Fig.~\ref{fig:FSC}
as a fixed-point form of the distribution function.

Figure~\ref{fig:FSC} clearly shows that the distribution function of
the random FM-AF chains has a quite different form from that of the
random AF chains.
It is essentially zero for $x/g(r)\gtrsim-1$ and increases approximately
linearly at $x/g(r)\lesssim-1$.
The weight function $W(x;r)$ representing the contribution to the mean
correlation is also shown in Fig.~\ref{fig:FSC}.
In contrast to the random AF case, $W(x;r)$ has most of its
weight in the region where $D(x;r)$ is not negligible.
This feature highlights the different nature of the spin correlations
in the random FM-AF chains.
As shown in the RSRG analysis, many spins in the random FM-AF chains
correlate and form a large effective spin.
This suggests that a spin is correlated with many other spins that
belong to the same large effective spin, and, therefore, the mean value
of the spin correlation function is not at all dominated by
the ^^ ^^ rare'' events that two spins, far apart from each other, form a
singlet pair.

From the observation that $D(x;r)/f(r)$ of the random FM-AF chains is
negligible for $x/g(r)>-A$ ($A\approx1$ in Fig.~\ref{fig:FSC}) and
has an approximately linear dependence for $x/g(r)<-A$ until it takes
a maximum value, we can also draw a conclusion that
$\ln\langle\langle C(r)\rangle\rangle$ and
$\langle\langle\ln C(r)\rangle\rangle$ should have a similar
dependence on $r$, in agreement with Figs.~\ref{fig:FMC} and
\ref{fig:FMlnC}.
Let us assume that the scaling function has
the form
\begin{equation}
F(y) = \left\{
 \begin{array}{@{\,}ll}
 -k ( y + A) \quad &  y \le -A,  \\
 0             &  y >   -A,
 \end{array}
 \right.
\label{eq:F}
\end{equation}
where $y=x/g(r)$, $k$ and $A$ are positive constants.
Since $g(r)\to\infty$ in the limit $r\to\infty$, the mean correlation
$\langle \langle C(r) \rangle \rangle$ is dominated by the
contribution from the region of small $|x/g(r)|$.
Thus, we are allowed to use Eq.~(\ref{eq:F}) for calculating
$\langle \langle C(r) \rangle \rangle$, even though this form
is not correct for large $|x/g(r)|$.
Using Eqs.~(\ref{eq:scl}),~(\ref{eq:scl1}) and ~(\ref{eq:F}), we
calculate the mean correlation:
\[
\langle \langle C(r) \rangle \rangle
= \int e^x D(x;r) dx
= -k \int^{-Ag(r)}_{-\infty}e^x\left(\frac{x}{g(r)}+A\right)dx
= \frac{k}{[g(r)]^2} e^{-A g(r)},
\]
yielding
\begin{equation}
\ln \langle \langle C(r) \rangle \rangle
= -A g(r) + {\cal O}\biglb( \ln g(r)\bigrb)
= Ac \langle\langle \ln C(r) \rangle\rangle
 + {\cal O}\biglb(\ln|\langle\langle\ln C(r)\rangle\rangle|\bigrb)
\label{eq:ln=ln}
\end{equation}
in the limit $r \to \infty$.
This result is in agreement with our observation that the curves in
Figs.~\ref{fig:FMC} and \ref{fig:FMlnC} look very similar.
It is also consistent with the fact that $W(x;r)$ in Fig.~\ref{fig:FSC}
stays almost in the same region of $x/g(r)$ with increasing $r$.
Equation (\ref{eq:ln=ln}) is not changed qualitatively even when $F(y)$
has an algebraic form $F(y)\propto[-(y+A)]^\alpha$, as far as $A>0$.
Note that Eq.~(\ref{eq:ln=ln}) does not hold in the random AF chains,
for which we should take $A=0$ and $\alpha=3$.

\section{Conclusion}
\label{sec:conclusion}

\begin{figure}
\epsfxsize=11cm
\epsffile{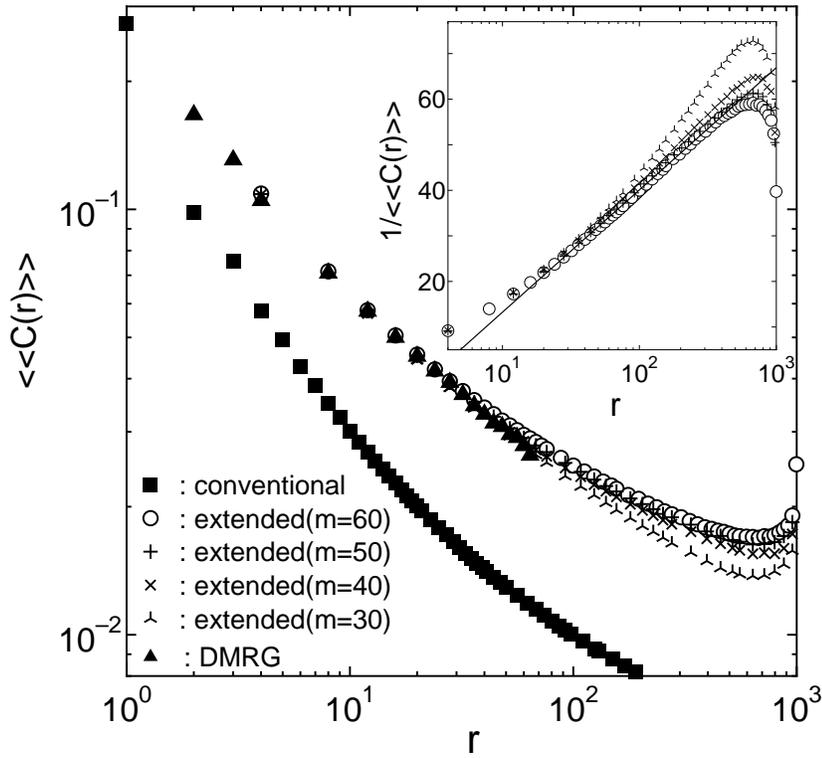}
   \caption{$\langle \langle C(r) \rangle \rangle$ of
            random FM-AF chains as a function of $r$.
            The statistical errors of the data
            are smaller than the size of symbols.
            Inset: $1/\langle\langle C(r)\rangle\rangle$ versus $r$.}
  \label{fig:FMC}
\end{figure}

\begin{figure}
\epsfxsize=11cm
\epsffile{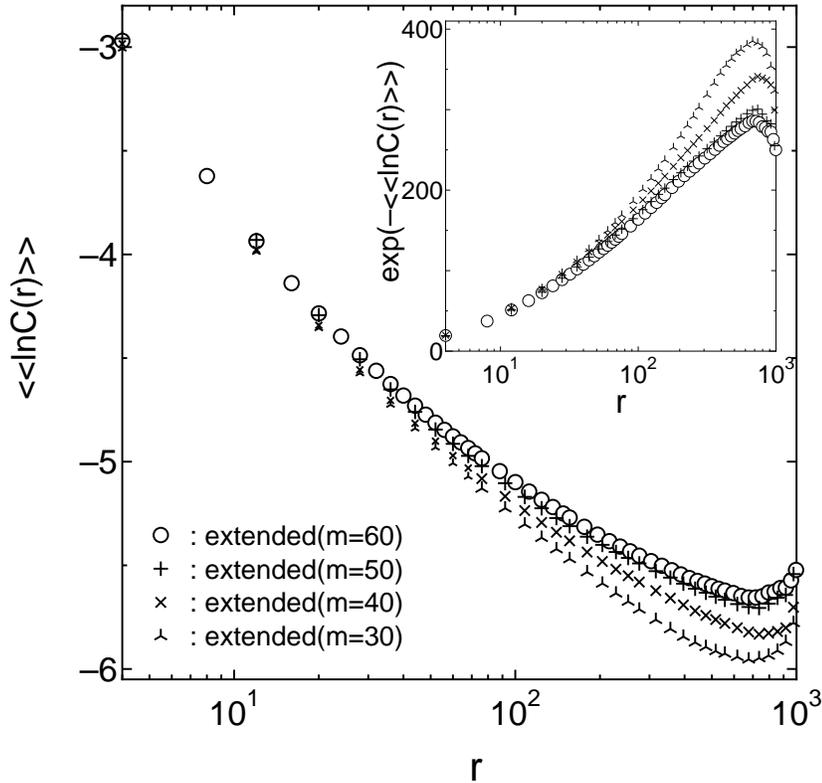}
   \caption{$\langle \langle \ln C(r) \rangle \rangle$ of
            the random FM-AF case as a function of $r$.
            The statistical errors of the data
            for $r < 500$ are smaller than the size of symbols.
            Inset: $\exp[-\langle\langle\ln C(r)\rangle\rangle]$ versus
$r$.}
  \label{fig:FMlnC}
\end{figure}

We have studied the spin correlations in spin-$\frac{1}{2}$
random Heisenberg chains for both the random AF and the random FM-AF case.
One of the most powerful methods for the study of random spin chains is
the real-space RG in which the spin degrees of freedom are decimated
by integrating out the strongest bonds successively.
In order to calculate the two-spin correlation functions between original
$S=\frac{1}{2}$ spins, we modified the algorithm of the conventional RSRG
using the Wigner-Eckart theorem.
We also extended the RSRG scheme by keeping the local excited
multiplets as block states to keep track of quantum effects.
We demonstrated that the extended RSRG algorithm is very powerful, in
the sense that it can be applied to rather large systems, and provides
accurate numerical data that are in excellent agreement with those of
the DMRG method.
The numerical data of the mean and typical correlations for the random
AF chains verified Fisher's prediction
[Eqs.~(\ref{eq:FS1}) and (\ref{eq:FS2})] for the Heisenberg case, which
cannot be mapped to noninteracting fermions.
For the random FM-AF chains, we found that the mean correlation,
$\langle \langle C(r) \rangle \rangle$, decays very slowly with the
logarithmic $r$-dependence.
We, therefore, conclude that the generalized staggered spin correlation
in the random FM-AF chains has no LRO, although it shows very
long-range correlations decaying slower than power-law behavior.
Investigating the distribution of the logarithm
of the spin correlation functions, we also clarified
the different nature of the ground states of the random AF and FM-AF chains.
In both cases, our data of distributions
satisfy the scaling hypothesis, Eq.~(\ref{eq:scl}).
Analyzing the form of the scaling function,
we found that the ^^ ^^ rare spin pairs", which are correlated
much stronger than typical ones, dominate the mean correlation
$\langle \langle C(r) \rangle \rangle$ in the random AF case.
It is also shown that the $r$-dependence of the mean correlations in
this case is quite different from that of the typical ones.
These features strongly support the fact that the ^^ ^^ random singlet
phase" is realized in the ground state of the random AF Heisenberg
chains.
On the other hand, the scaling form of the distribution functions
for the random FM-AF chains suggests that such ^^ ^^ rare spin pairs"
do not play an important role in these chains.
We also deduced that $\ln \langle \langle C(r) \rangle \rangle$ and
$\langle \langle \ln C(r) \rangle \rangle$ have very similar
$r$-dependence, using the scaling hypothesis and the specific feature
of the scaling function we obtained numerically.
This result is non-trivial and consistent with our numerical data
(Figs.~\ref{fig:FMC} and \ref{fig:FMlnC}).

\acknowledgements

HT was supported by the Research Fellowship of the Japan Society for
the Promotion of Science for Young Scientists.
The work of AF and MS was in part supported by a Grant-in-Aid from the
Ministry of Science, Education and Culture of Japan.
Numerical calculations were carried out at the Yukawa Institute
Computing Facility.

\newpage
\narrowtext

\begin{table}
\begin{tabular}{lr}
RG scheme & number of samples \\
\hline
conventional RSRG        &  200  \\
extended RSRG ($m = 10$) & 2460  \\
extended RSRG ($m = 20$) & 1878  \\
extended RSRG ($m = 30$) & 1899  \\
DMRG                     & 1887  \\
\end{tabular}
\caption{The number of samples simulated to take the random averages
in the random AF chains.}
\label{tb1}
\end{table}%

\begin{table}
\begin{tabular}{lr}
RG scheme & number of samples \\
\hline
conventional RSRG        & 150 \\
extended RSRG ($m = 30$) & 1319 \\
extended RSRG ($m = 40$) & 1446 \\
extended RSRG ($m = 50$) & 1890 \\
extended RSRG ($m = 60$) & 1557 \\
DMRG                     & 204 \\
\end{tabular}
\caption{The number of samples simulated to take the random averages
in the random FM-AF chains.}
\label{tb2}
\end{table}

\widetext

\begin{figure}
\epsfxsize=11cm
\epsffile{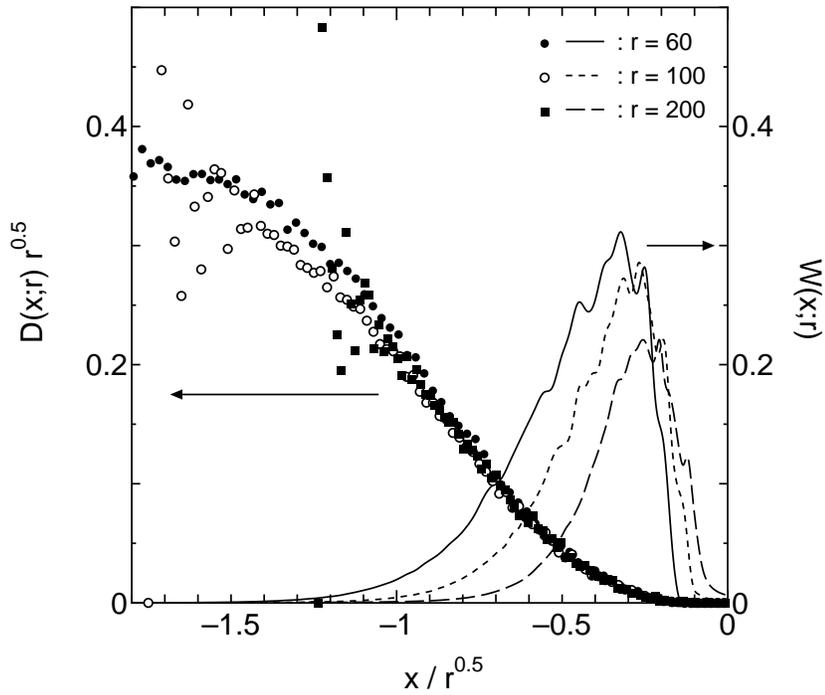}
   \caption{Probability distributions of the correlation functions
    in the random AF case for $r = 60,100,200$.
    $W(x;r)$ is also plotted for $r = 60,100,200$ with
    a solid, dotted, dashed curve, respectively.}
  \label{fig:ASC}
\end{figure}%


\begin{references}
\bibitem{stretched}
  We also tried the stretched exponential form,
  $\exp\{-\tilde a[\ln(r/\tilde r_0)]^\gamma\}$, with which we could
  fit the numerical data as well as with Eq.~(\ref{eq:log}).
  Here $\tilde a$, $\tilde r_0$, and $\gamma$ are constants.
\bibitem{note}
  Some data points at $x/g(r)<-1$ are off the scaling curve in
  Fig.~\ref{fig:ASC}.
  We estimate that the numerical error in $C(r)$ in our calculation is
  of order $10^{-8}$.
  The scattered data are those with small correlations of order
  $10^{-7}$.  Thus the failure of the data collapse at $x/g(r)<-1$
  does not mean the violation of the scaling.
\end{references}
\end{document}